\documentclass[conference]{IEEEtran}
\IEEEoverridecommandlockouts
\usepackage{cite}
\usepackage{amsmath,amssymb,amsfonts}
\usepackage{algorithmic}
\usepackage{graphicx}
\usepackage{textcomp}
\usepackage{xcolor}
\usepackage{subfigure}
\def\BibTeX{{\rm B\kern-.05em{\sc i\kern-.025em b}\kern-.08em
    T\kern-.1667em\lower.7ex\hbox{E}\kern-.125emX}}

\IEEEoverridecommandlockouts
\IEEEpubid{\makebox[\columnwidth]{978-1-7281-9320-5/20/\$31.00~\copyright~2020 IEEE \hfill} \hspace{\columnsep}\makebox[\columnwidth]{ }}
\begin{document}

\title{Learned Variable-Rate Multi-Frequency Image Compression using Modulated Generalized Octave Convolution\\
}

\author{\IEEEauthorblockN{Jianping Lin}
\IEEEauthorblockA{\textit{University of Science and Technology of China} \\
\textit{and Simon Fraser University, Canada}\\
 ljp105@mail.ustc.edu.cn}
\and
\IEEEauthorblockN{Mohammad Akbari}
\IEEEauthorblockA{\textit{Simon Fraser University, Canada} \\
 akbari@sfu.ca}
\and
\IEEEauthorblockN{Haisheng Fu}
\IEEEauthorblockA{\textit{Xi'an Jiaotong University, China} \\
 fhs381408020105@126.com}
\and
\IEEEauthorblockN{Qian Zhang}
\IEEEauthorblockA{\textit{Xi'an Jiaotong University, China} \\
 zhangqian@stu.xjtu.edu.cn}
\and
\IEEEauthorblockN{Shang Wang}
\IEEEauthorblockA{\textit{Xi'an Jiaotong University, China} \\
 798173752@qq.com}
\and
\IEEEauthorblockN{Jie Liang}
\IEEEauthorblockA{\textit{Simon Fraser University, Canada} \\
  jiel@sfu.ca}
\and
\IEEEauthorblockN{Dong Liu}
\IEEEauthorblockA{\textit{University of Science and Technology of China} \\
  dongeliu@ustc.edu.cn}
\and
\IEEEauthorblockN{Feng Liang}
\IEEEauthorblockA{\textit{Xi'an Jiaotong University, China} \\
  fengliang@mail.xjtu.edu.cn}
\and
\IEEEauthorblockN{Guohe Zhang}
\IEEEauthorblockA{\textit{Xi'an Jiaotong University, China} \\
  zhangguohe@xjtu.edu.cn}
\and
\IEEEauthorblockN{Chengjie Tu}
\IEEEauthorblockA{\textit{Tencent Technologies, China} \\
  chengjietu@tencent.com}
}

\maketitle

\begin{abstract}
In this proposal, we design a learned multi-frequency image compression approach that uses generalized octave convolutions to factorize the latent representations into high-frequency (HF) and low-frequency (LF) components, and the LF components have lower resolution than HF components, which can improve the rate-distortion performance, similar to wavelet transform. Moreover, compared to the original octave convolution, the proposed generalized octave convolution (GoConv) and octave transposed-convolution (GoTConv) with internal activation layers preserve more spatial structure of the information, and enable more effective filtering between the HF and LF components, which further improve the performance. In addition, we develop a variable-rate scheme using the Lagrangian parameter to modulate all the internal feature maps in the auto-encoder, which allows the scheme to achieve the large bitrate range of the JPEG AI with only three models. Experiments show that the proposed scheme achieves much better Y MS-SSIM than VVC. In terms of YUV PSNR, our scheme is very similar to HEVC.
\end{abstract}


\section{Method}

\subsection{Overview of the Encoding/Decoding Architectures}

Recently, deep learning-based image compression has shown the potential to outperform standard codecs such as JPEG2000 \cite{skodras2001jpeg}, the H.265/HEVC Intra-based BPG image codec \cite{sullivan2012overview}, and the intra coding of the upcoming versatile video coding (VVC) standard  \cite{vvc-vtm}. In particular, the scheme in  \cite{akbari2020generalized} even achieves better PSNR than VVC for the Kodak dataset, based on the recently developed octave convolution \cite{chen2019drop} and the entropy coding scheme in \cite{Minnen18}, where hyperprior and autoregressive models are jointly utilized to effectively capture the spatial dependencies in the latent representations. However, the autoregressive model has a disadvantage of high decoding complexity.

In this proposal, a simplified scheme of \cite{akbari2020generalized} is developed. Since our model is based on the float32 precision, the original 8-bit image in RGB format is first divided by 256 for normalization. The R-D tradeoff Lagrangian parameter \(\lambda\) is also input into the encoder to control the bitrate, similar to \cite{choi2019variable}. The decoder requires the same \(\lambda\) value to do the  reconstruction, so we first quantize the \(\lambda\) value into int32 format and include it in the bitstream. At the decoder side, we first dequantize the \(\lambda\) value back to float32 format, and then use it to decode the received bitstream to reconstruct the decoded 8-bit image in RGB format.

\subsection{Modulated Generalized Octave Convolution}

Our deep learning network is generalized from the recently developed octave convolution \cite{chen2019drop}, where the feature maps are factorized into high-frequency (HF) and low-frequency (LF) components, and the LF components have lower resolution than the HF components, which save both memory and computation. This framework is very suitable for image compression, similar to the idea of the wavelet transform, which has been used in JPEG2000 \cite{skodras2001jpeg}.

\begin{figure*}
  \centering
  \subfigure[]{
    %\label{fig:subfig:a} %% label for first subfigure
    \includegraphics[width=.42 \linewidth]{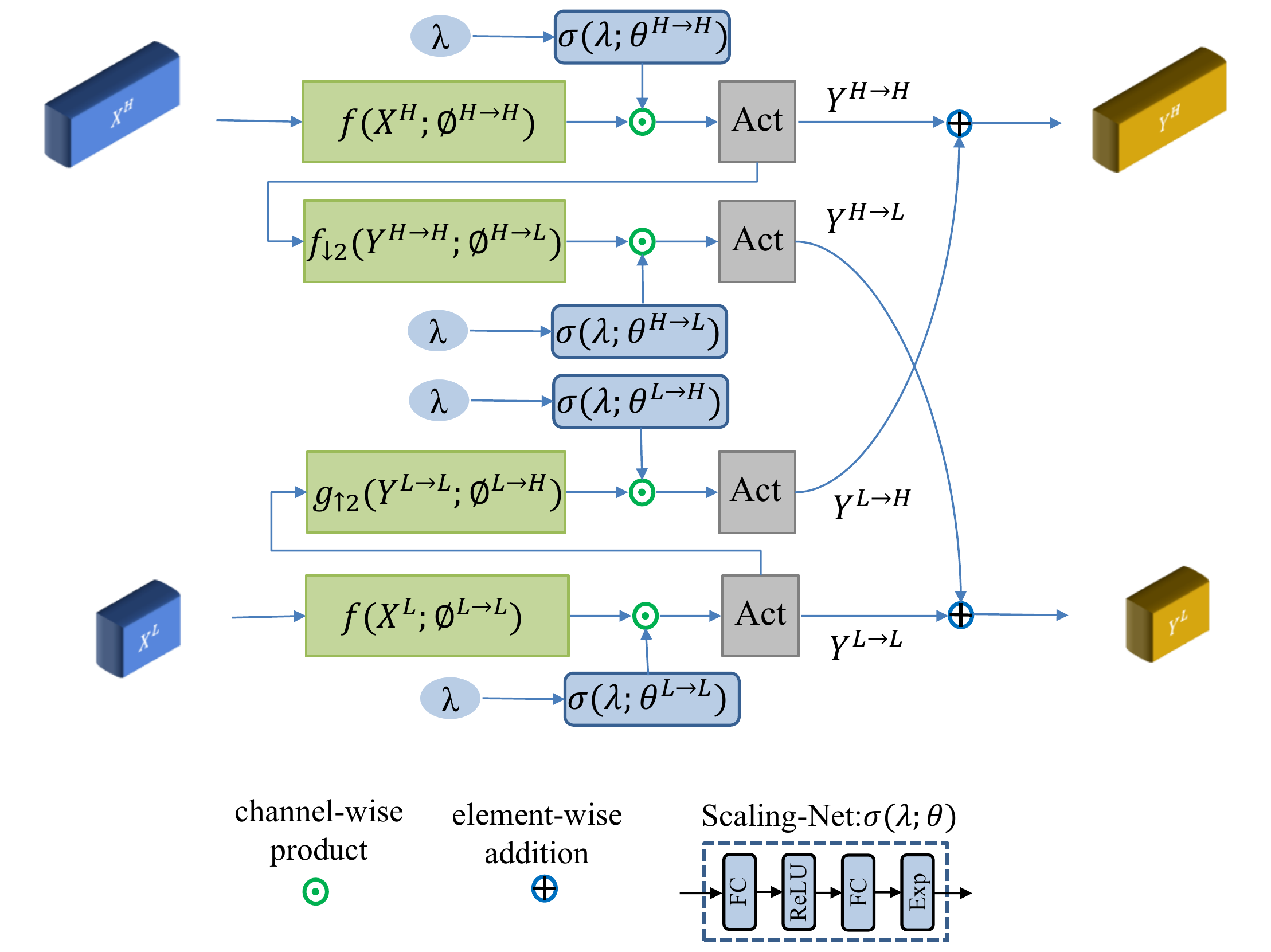}
    }
  %\hspace{1.0in}
  \subfigure[]{
    %\label{fig:subfig:b} %% label for second subfigure
    \includegraphics[width=.45\linewidth]{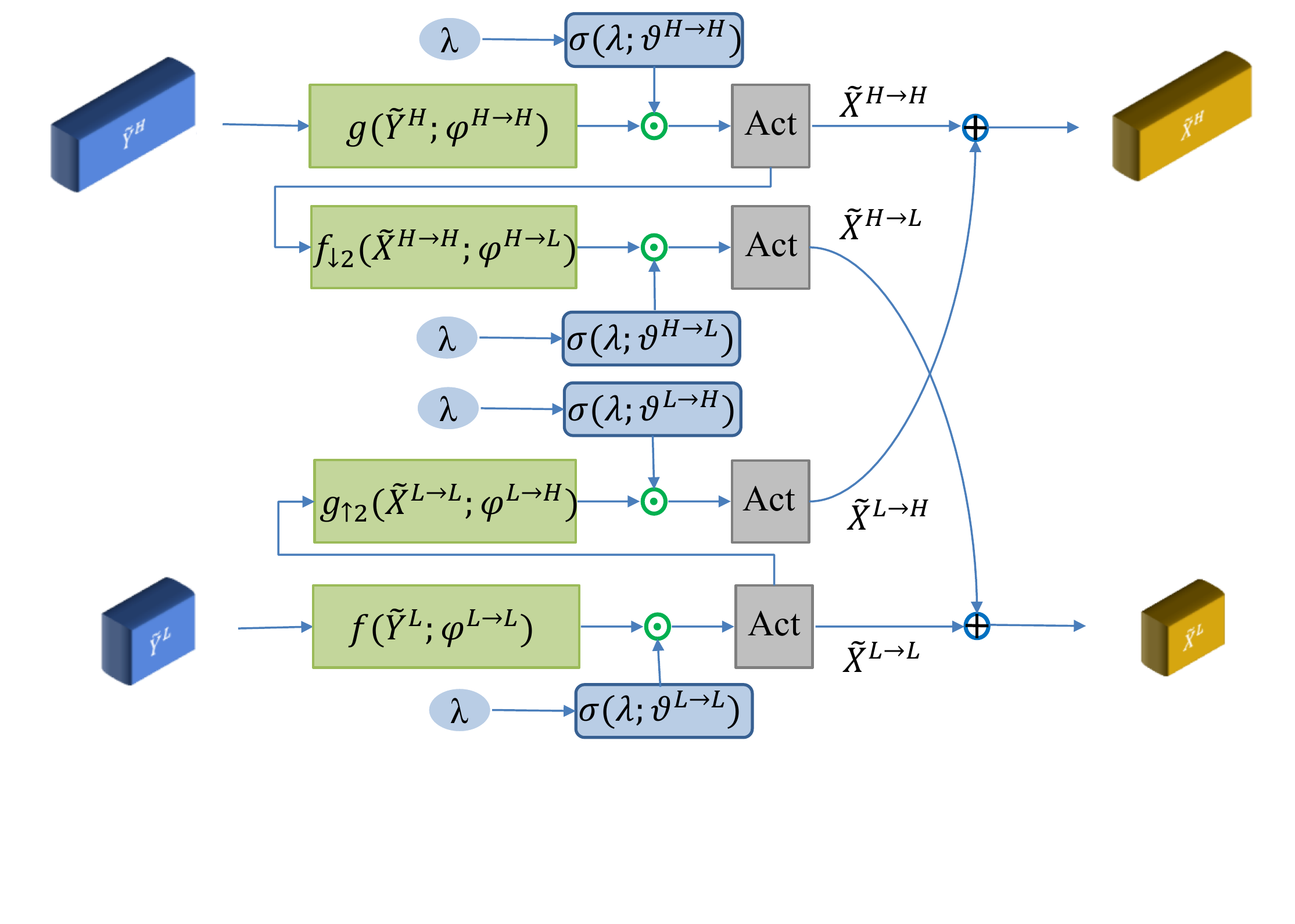}}
    %\hspace{1.0in}
  \caption{Architecture of the proposed generalized octave convolution modulated by \(\lambda\) (M-GoConv) (a) and transposed-convolution modulated by \(\lambda\) (M-GoTConv) (b). \textit{Act}: the activation layer; $f$: regular vanilla convolution; $g$: regular transposed-convolution; $f_{\downarrow2}$: regular convolution with stride 2; $g_{\uparrow2}$: regular transposed-convolution with stride 2.}
  \label{fig:M-GoConv} %% label for entire figure
\vspace{-0.15cm}
\end{figure*}

Fig.~\ref{fig:M-GoConv} shows the architectures of the proposed \(\lambda\)-Modulated Generalized Octave Convolution (M-GoConv) and \(\lambda\)-Modulated Generalized Octave Transposed Convolution (M-GoTConv).
The output HF and LF frequency feature maps \(Y^{H}\) and \(Y^{L}\) in M-GoConv are formulated as follows:
\begin{equation}
\begin{split}
       Y^{H} = Y^{H\rightarrow H}+Act\left(g_{\uparrow2}(Y^{L\rightarrow L};\phi^{L\rightarrow H})\odot\sigma(\lambda;\theta^{L\rightarrow H})\right), \\
       Y^{L} = Y^{L\rightarrow L}+Act\left(f_{\downarrow2}(Y^{H\rightarrow H};\phi^{H\rightarrow L})\odot\sigma(\lambda;\theta^{H\rightarrow L})\right), \\
      \text{ with }
      Y^{H\rightarrow H} =Act\left(f(X^H;\phi^{H\rightarrow H})\odot\sigma(\lambda;\theta^{H\rightarrow H})\right),\\
      Y^{L\rightarrow L}=Act\left(f(X^L;\phi^{L\rightarrow L})\odot\sigma(\lambda;\theta^{L\rightarrow L})\right),
\end{split}
\label{equation:M-GoConv}
\end{equation}
where \(X^H\) and \(X^L\) are the input HF and LF feature maps. \(f_{\downarrow2}\) and \(g_{\uparrow2}\) are respectively Vanilla convolution and transposed-convolution operations with stride of 2. \(\phi\) and \(\theta\) are the parameters of convolutions. \(\sigma\) stands for the function of the scaling-network (Scaling-net) used to map the scalar value of \(\lambda\) into a vector to channel-wisely scale the feature map after the convolution operation. \(\odot\) stands for the channel-wise product operation. $Act$ presents the activation layer which can be any function like GDN or Leaky ReLU.

Similarly, the output HF and LF feature maps \(\Tilde{X}^{H}\) and \(\Tilde{X}^{L}\) in M-GoTConv are obtained as follows:
\begin{equation}
\begin{split}
       \Tilde{X}^{H} = \Tilde{X}^{H\rightarrow H}+Act\left(g_{\uparrow2}(\Tilde{X}^{L\rightarrow L};\varphi^{L\rightarrow H})\odot\sigma(\lambda;\vartheta^{L\rightarrow H})\right), \\
       \Tilde{X}^{L} = \Tilde{X}^{L\rightarrow L}+Act\left(f_{\downarrow2}(\Tilde{X}^{H\rightarrow H};\varphi^{H\rightarrow L})\odot\sigma(\lambda;\vartheta^{H\rightarrow L})\right), \\
      \text{ with }      \Tilde{X}^{H\rightarrow H} = Act\left(g(\Tilde{Y}^H;\varphi^{H\rightarrow H})\odot\sigma(\lambda;\vartheta^{H\rightarrow H})\right),\\
      \Tilde{X}^{L\rightarrow L} = Act\left(g(\Tilde{Y}^L;\varphi^{L\rightarrow L})\odot\sigma(\lambda;\vartheta^{L\rightarrow L})\right),
\end{split}
\label{equation:M-GoTConv}
\end{equation}
where \(\Tilde{Y}^H\) and \(\Tilde{Y}^L\) are the input HF and LF feature maps. $\Tilde{Y}^H, \Tilde{X}^H \in \mathbb{R}^{h\times w \times(1-\alpha)c}$ and $\Tilde{Y}^L, \Tilde{X}^L \in \mathbb{R}^{\frac{h}{2} \times \frac{w}{2} \times \alpha c}$, where \(\alpha\) is the portion of LF components in all feature maps. Same as our previous work \cite{akbari2020generalized}, we set \(\alpha\) as 0.5 in our experiments.

\subsection{Proposed Variable-rate Multi-Frequency Model}

\begin{figure*}
 \centering
  \centerline{\includegraphics[width=0.8 \textwidth]{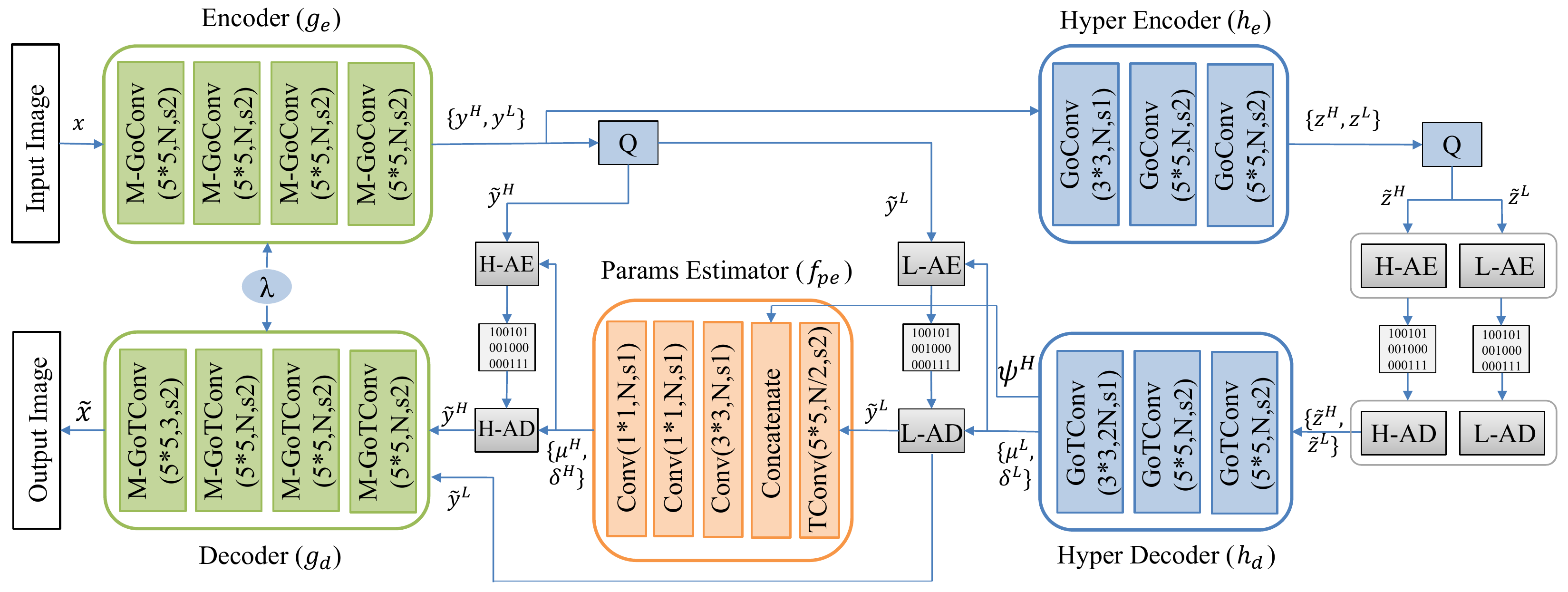}}
\caption{Overall framework of the proposed image compression model. \textbf{H-AE} and \textbf{H-AD}: arithmetic encoder and decoder for high frequency latents. \textbf{L-AE} and \textbf{L-AD}: arithmetic encoder and decoder for low frequency latents. \textbf{Q}: represents the additive uniform noise for training, or uniform quantizer for the test.}
%\vskip -10pt
\label{fig:framework}
\vspace{-0.20cm}
\end{figure*}

Fig.~\ref{fig:framework} illustrates the overall architecture of the proposed variable-rate multi-frequency image compression framework. Similar to \cite{balle2018variational}, our architecture is composed of two parts: the core autoencoder and the entropy sub-network. The core autoencoder is used to learn a quantized latent vector of the input image, while the entropy sub-network is responsible for learning a probabilistic model over the quantized latent representations, which is utilized for entropy coding. The time-consuming autoregressive model part in \cite{Minnen18,akbari2020generalized} is not used in this proposal, in order to speed up the encoding and decoding.

We have made several improvements to the scheme in \cite{balle2018variational}. In order to handle multi-frequency entropy coding, all vanilla convolutions in the core encoder (decoder) are replaced by the proposed M-GoConv (M-GoTConv), and all vanilla transposed-convolutions in the hyper encoder (decoder) are replaced by GoConv (GoTConv). Here, the GoConv and GoTConv are the networks by removing the Scaling-Nets from M-GoConv and M-GoTConv, respectively. In \cite{balle2018variational}, each convolution/transposed-convolution is accompanied by an activation layer (e.g., GDN/IGDN or Leaky ReLU). In our architecture, we move these layers into the GoConv and GoTConv architectures and directly apply them to the inter- and intra-frequency components. GDN/IGDN transforms are respectively used for the M-GoConv and M-GoTConv employed in the core autoencoder, while Leaky ReLU is utilized for the hyper autoencoder.

To further exploit the correlation between the quantized LF and HF latent representations $\Tilde{y}^L$ and $\Tilde{y}^H$, we first encode the LF components $\Tilde{y}^L$ and then use the decoded $\Tilde{y}^L$ together with the HF hyper parameters $\psi^H$ to estimate the distribution of HF components $\Tilde{y}^H$. This is an improvement over \cite{akbari2020generalized}.

Let $x \in \mathbb{R}^{h\times w\times 3}$ be the input image and \(\lambda\) is the input R-D tradeoff parameter, the multi-frequency latent representations are denoted by $\{y^{H},y^{L}\}$ where $y^{H} \in \mathbb{R}^{\frac{h}{16} \times \frac{w}{16} \times (1-\alpha)N}$ and $y^{L} \in \mathbb{R}^{\frac{h}{32} \times \frac{w}{32} \times \alpha N}$ are generated using the parametric deep encoder (i.e., analysis transform) $g_e$ represented as:
\begin{equation}
\{y^{H},y^{L}\}=g_e(x,\lambda;\theta_{ge}),
\end{equation}
where $\theta_{ge}$ is the parameter vector to be optimized. $N$ denotes the total number of output channels in $g_e$, which is divided into $(1-\alpha)N$ HF channels and $\alpha N$ LF channels, which have half of the spatial resolution of the HF channels.

At the decoder side,
%given the quantized latents $\{\Tilde{y}^{H},\Tilde{y}^{L}\}$,
the parametric decoder (i.e., synthesis transform) $g_d$ with the parameter vector $\theta_{gd}$ reconstructs the image $\Tilde{x} \in \mathbb{R}^{h\times w\times 3}$ as follows:
\begin{equation}
\begin{split}
    \Tilde{x}=g_d\left(\{\Tilde{y}^{H},\Tilde{y}^{L}\},\lambda;\theta_{gd}\right)
    \text{  with  }
    \{\Tilde{y}^{H},\Tilde{y}^{L}\} &= Q\left(\{y^H,y^L\}\right),
\end{split}
\end{equation}
where $Q$ represents the addition of uniform noise to the latent representations during training, or uniform quantization (i.e., rounding function in this proposal) and arithmetic coding/decoding of the latents during the test. As illustrated in Fig.~\ref{fig:framework}, the quantized HF and LF latents $\Tilde{y}^{H}$ and $\Tilde{y}^{L}$ are entropy-coded using two separate arithmetic encoder and decoder.

The hyper autoencoder learns to represent side information used to estimate the distribution of the quantized HF and LF latents $\Tilde{y}^{H}$ and $\Tilde{y}^{L}$. The spatial dependencies of $\{\Tilde{y}^{H},\Tilde{y}^{L}\}$ are then captured into the multi-frequency hyper latent representations $\{z^{H},z^{L}\}$ using the parametric hyper encoder $h_e$ (with the parameter vector $\theta_{he}$) defined as:
\begin{equation}
\begin{split}
    \{z^{H},z^{L}\}=h_e\left(\{{y}^{H},{y}^{L}\};\theta_{he}\right).
\end{split}
\end{equation}

The quantized hyper latents are also part of the generated bitstream that is required to be entropy-coded and transmitted. Similar to the core latents, two separate arithmetic coders are used for the quantized HF and LF components $\Tilde{z}^H$ and $\Tilde{z}^L$. Given the quantized hyper latents, the mean and scale parameters of the conditional Gaussian entropy model for the quantized LF latents $\Tilde{y}^{L}$ and the side information used for the entropy model estimation of the quantized HF latents $\Tilde{y}^{H}$ is reconstructed using the hyper decoder $h_d$ (with the parameter vector $\theta_{hd}$) formulated as:
\begin{equation}
\begin{split}
    \{\mu^L, \delta^L, \psi^H\}=h_d\left(\{\Tilde{z}^H,\Tilde{z}^L\};\theta_{hd}\right)\\
    \text{  with  }
    \{\Tilde{z}^{H},\Tilde{z}^{L}\} = Q\left(\{z^H,z^L\}\right).
\end{split}
\end{equation}
where $\mu^L$ and $\delta^L$ are the parameters for entropy modelling of the LF information, and $\psi^H$ is the side information for the HF information.

As shown in Fig.~\ref{fig:framework}, to estimate the mean and scale parameters of the conditional Gaussian entropy model for the quantized HF latents $\Tilde{y}^{H}$, the information from both entropy decoded $\Tilde{y}^{L}$ and the side information $\psi^H$ is combined by another network, denoted by $f_{pe}$ (with the parameter vector $\theta_{ep}$), represented as follows:
\begin{equation}
    \{\mu^H, \delta^H\}=
    f_{pe}\left(\Tilde{y}^{L}, \psi^H;\theta_{ep}\right),
\label{equation:parameters estimator}
\end{equation}
where $\mu^H$ and $\delta^H$ are the parameters for entropy modelling of the HF information.

\subsection{Methodology for Training}

\subsubsection{Dealing with Quantization}

As in \cite{balle2018variational}, we use the addition of uniform noise to the latent representations during training, or uniform quantization (i.e., rounding function in this proposal) and arithmetic coding/decoding of the latents during the test.

\subsubsection{Loss function}

\begin{table*}
\caption{The $\lambda$ sets used to train different variable-rate models.}
\label{lambda_set}
\center
\begin{tabular}{c|c|c}
\hline
Bitrate range &$\lambda$ set & Target \\
of Model      &     & bitrate (bpp) \\
\hline
Low        & 0.000001,0.000005,0.00001,0.00003,0.00007,0.0001,0.0003,0.0007,0.001,0.003,0.005                   &0.06,0.12	           \\
\hline
Middle       & 0.0001,0.0003,0.0007,0.001,0.003,0.005,0.007,0.01,0.03,0.05,0.07,0.1                   &0.25,0.5	        \\
\hline
High       & 0.0001,0.0003,0.0007,0.001,0.003,0.005,0.007,0.01,0.03,0.05,0.07,0.1,0.2,0.3,0.4,0.5                    &0.75,1.0,1.5,2.0	        \\
\hline
\end{tabular}
\vspace{-0.15cm}
\end{table*}

The objective function for training is composed of two terms: rate $R$, which is the expected length of the bitstream, and distortion $D$, which is the expected error between the input and reconstructed images. Our scheme takes the Lagrangian multiplier $\lambda$ as a conditioning input parameter and aims to produce the reconstructed image $\tilde{x}$ with varying rate and distortion depending on the conditioning value of $\lambda$. The R-D optimization problem is then defined as follows:
\begin{equation}
\begin{split}
    \mathcal{L}&=\sum_{\lambda \in \Lambda}(R+\lambda D)\\ \text{  with  }
    R&=R^H+R^L,
    D=\mathbb{E}_{x	\sim p_x}\left[d(x,\tilde{x})\right],
\end{split}
\label{equation:R-D}
\end{equation}
where $p_x$ is the unknown distribution of natural images and $D$ is the distortion between $x$ and $\tilde{x}$, which is measured by the weighted sum of RGB mean squared error ($RGB\_MSE$) and $1-Y\_MS$-$SSIM$, i.e., $0.9*(RGB\_MSE) + 0.1*(1-Y\_MS$-$SSIM)$, in our experiments. $\Lambda$ is a predefined $\lambda$ set for training. As suggested by \cite{choi2019variable}, we randomly select $\lambda$ from $\Lambda$ for each training sample to compute its individual R-D cost, and then we use the average R-D cost per batch as the loss for gradient descent. $R^H$ and $R^L$ are the rates corresponding to the HF and LF information (bitstreams) defined as follows:
\begin{equation}
\begin{split}
    R^H=\prod_{i}\left(\mathcal{N}(\mu^H_i,{\delta_i^{2}}^H)*\mathcal{U}(-\tfrac{1}{2},\tfrac{1}{2}) \right)(\Tilde{y}^H_i),\\
    +\prod_{j}\left(p_{{z^H_i}|\Theta^H_j}(\Theta^H_j)*\mathcal{U}(-\tfrac{1}{2},\tfrac{1}{2}) \right)(\Tilde{z}^H_j),\\
    R^L=\prod_{i}\left(\mathcal{N}(\mu^L_i,{\delta_i^{2}}^L)*\mathcal{U}(-\tfrac{1}{2},\tfrac{1}{2}) \right)(\Tilde{y}^L_i),\\
    +\prod_{j}\left(p_{{z^L_i}|\Theta^L_j}(\Theta^L_j)*\mathcal{U}(-\tfrac{1}{2},\tfrac{1}{2}) \right)(\Tilde{z}^L_j),\\
\end{split}
\label{equation:rate}
\end{equation}
where $\Theta^H$ and $\Theta^L$ denote the parameter vectors for the univariate distributions $p_{\Tilde{z}^H|\Theta^H}$ and $p_{\Tilde{z}^L|\Theta^L}$.

\subsubsection{Preparation of Training Data}

We crop image patches of size 256x256 pixels from the original images of the JPEG-AI training set to train the models. The batch size is set as 12. No other data augmentations (e.g. down-sampling, flipping etc.) have been applied in our training data.

\subsubsection{Training Procedure}

We set $\alpha=0.5$ so that 50\% of the latent representations is assigned to the LF part with half spatial resolution. We set the number of filters (the N value in Fig.~\ref{fig:framework}) as 448. To cover the large bitrate range of the JPEG AI CfE, we train three different variable-rate models using different $\lambda$ sets, which are summarized in table~\ref{lambda_set}. Our models are trained by 2,000,000 steps with a fixed learning rate of 0.0004. Our models are implemented by TensorFlow, and trained on a single Titan Xp/GTX 1080Ti GPU.

\subsection{Model Storage Analysis}

To cover the specified large bitrate range of the CfE, we train three different variable-rate models. All of them are based on the precision of float32. The size of each encoder and decoder model  with N = 448 is about 120MB.

\section{Experimental Results}

\begin{figure*}
  \centering
  \subfigure[]{
    %\label{fig:subfig:a} %% label for first subfigure
    \includegraphics[width=2.65in]{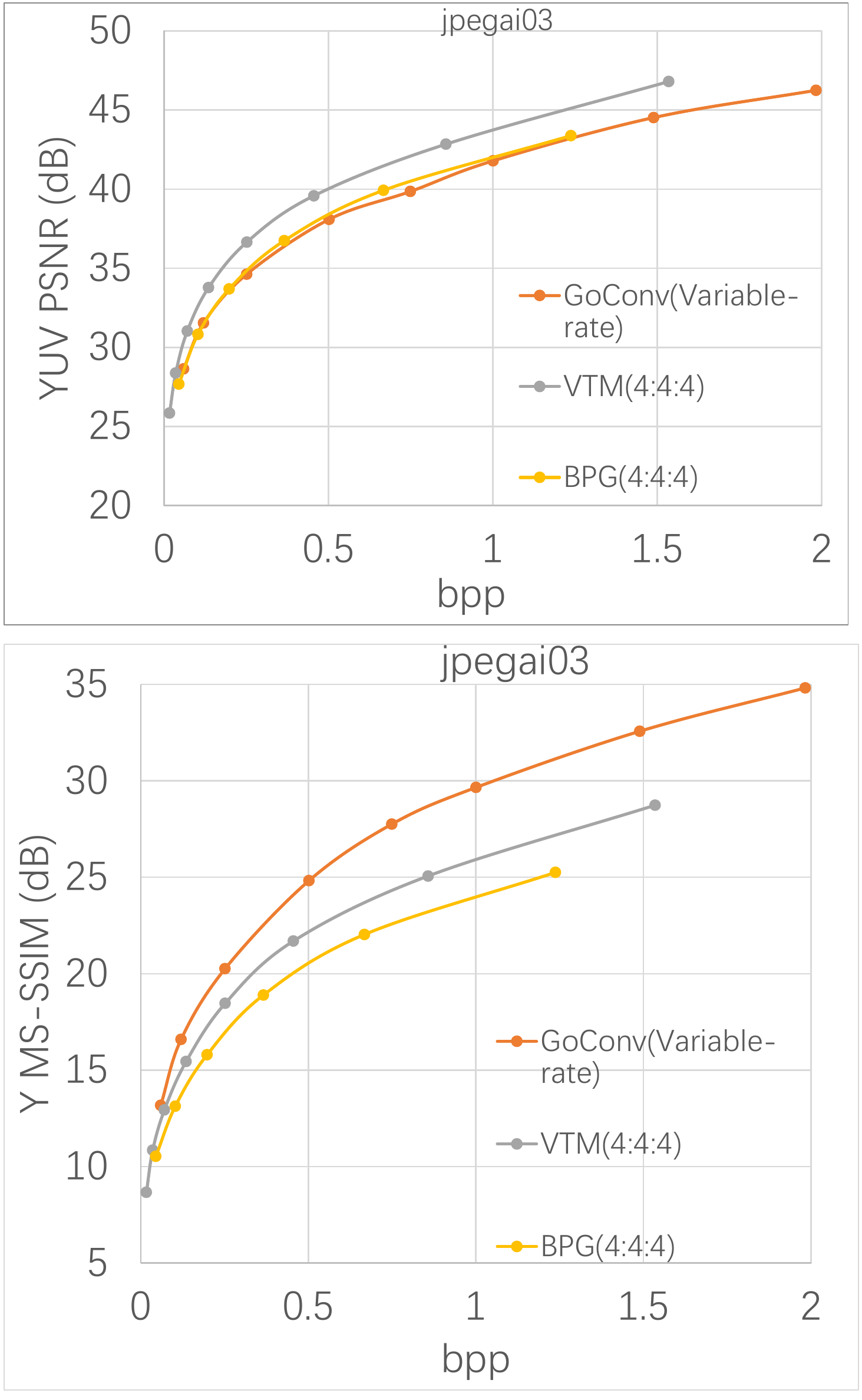}}
  %\hspace{1.0in}
  \subfigure[]{
    %\label{fig:subfig:b} %% label for second subfigure
    \includegraphics[width=2.65in]{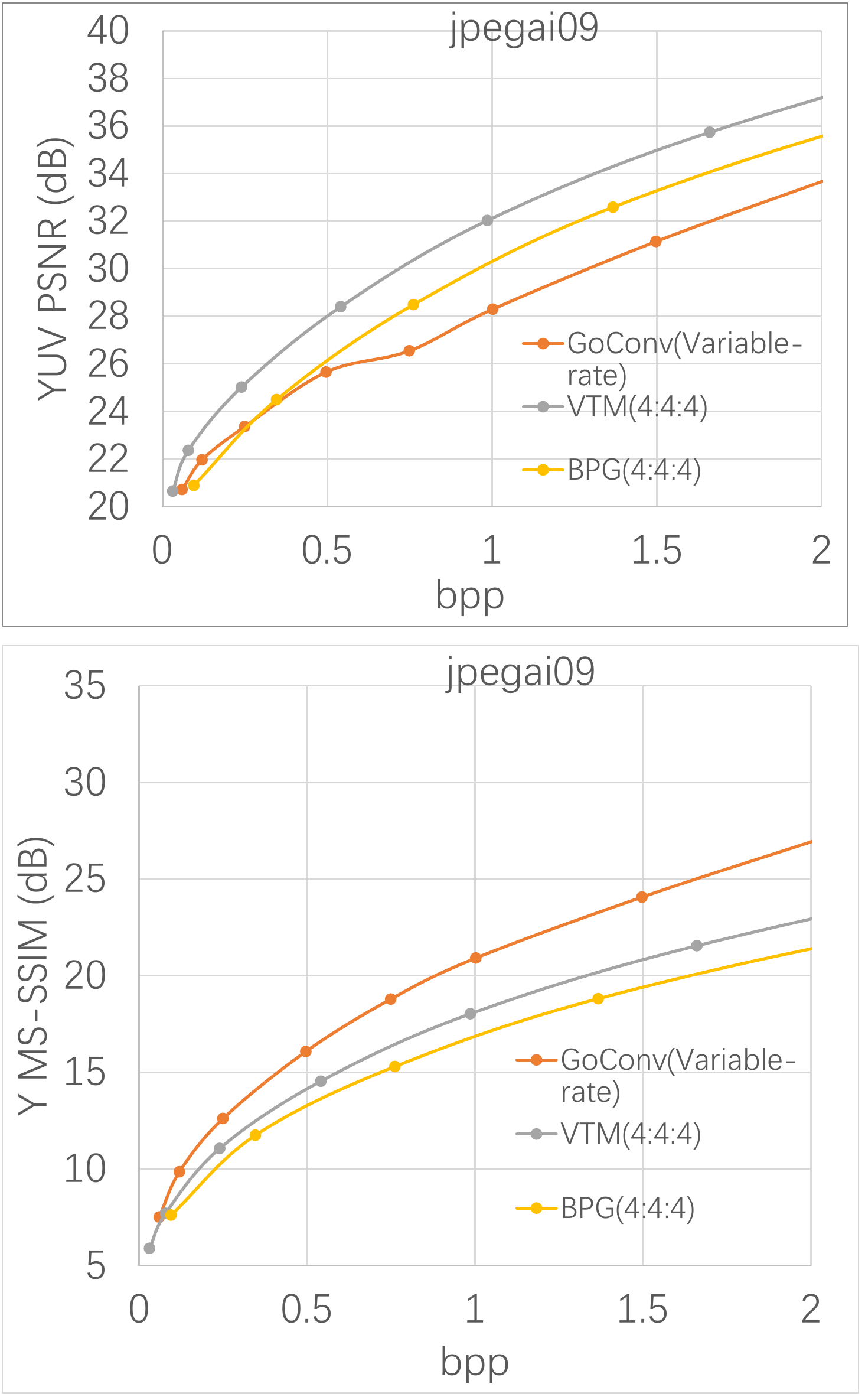}}
    %\hspace{1.0in}
  \subfigure[]{
    %\label{fig:subfig:c} %% label for second subfigure
    \includegraphics[width=2.65in]{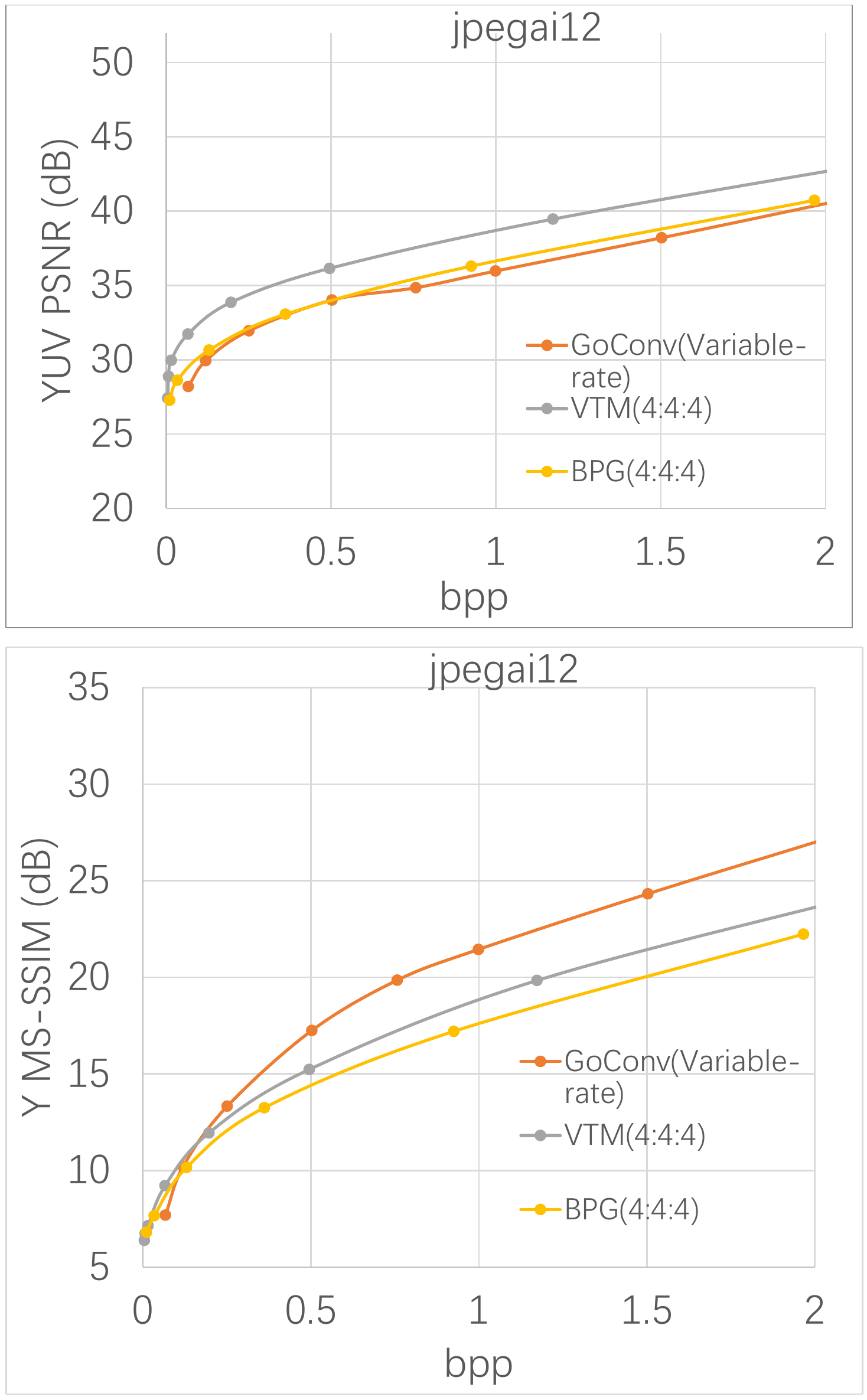}}
  \subfigure[]{
    %\label{fig:subfig:a} %% label for first subfigure
    \includegraphics[width=2.65in]{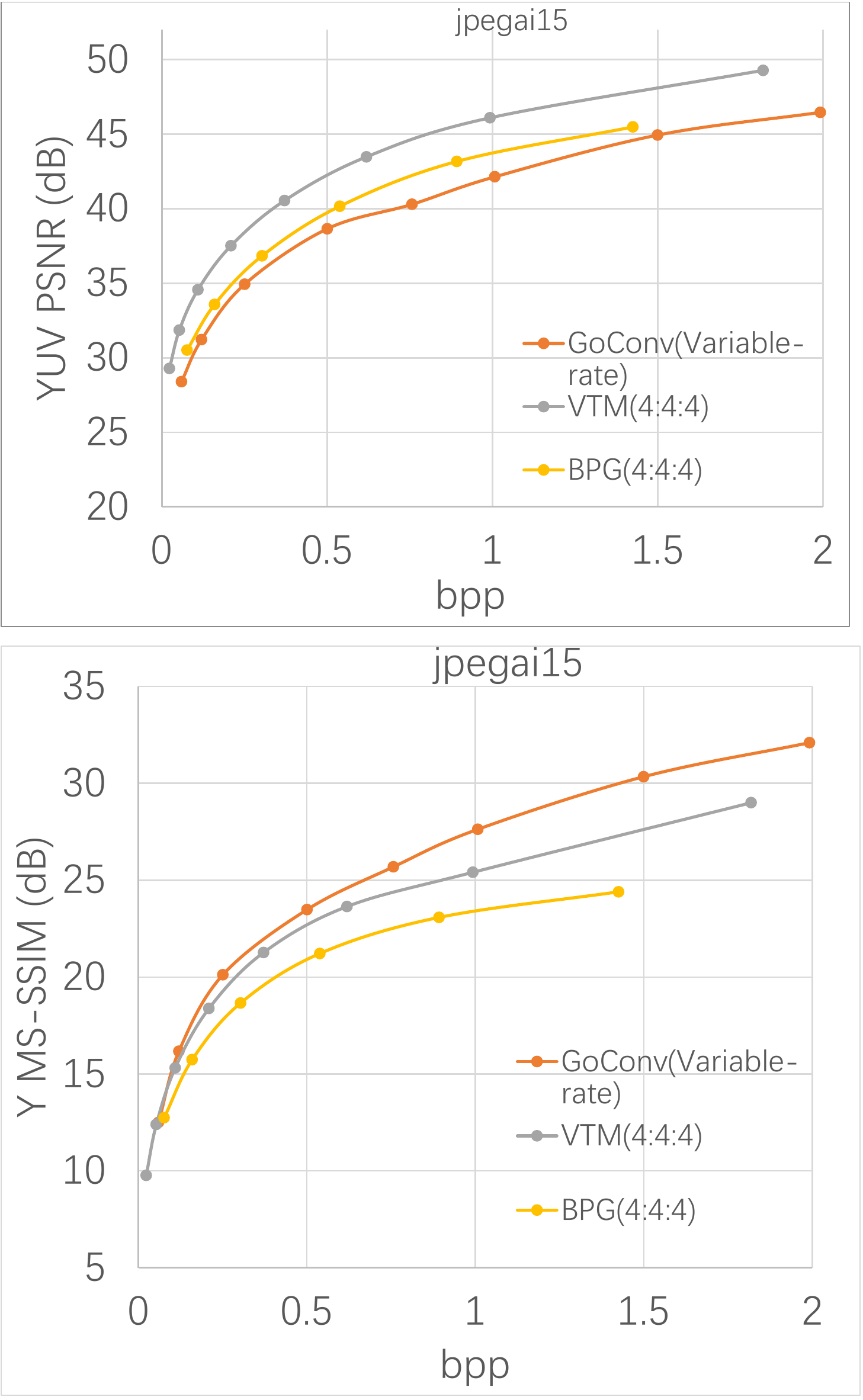}}
    %\hspace{1.0in}
  \caption{\textbf{Overall performance}. The compression results of the four specified test images in $YUV\_PSNR$ and $Y MS$-$SSIM (dB)$ using the proposed method (denoted by GoConv(Variable-rate)), BPG \cite{sullivan2012overview}, and VTM \cite{vvc-vtm}.}
  \label{fig_RD Curve} %% label for entire figure
\end{figure*}

The R-D performance and encoding/decoding time of four images in the JPEG AI testset are reported in Table \ref{test03} to Table \ref{time}. The distortion is measured by $YUV\_PSNR$ and $Y MS$-$SSIM$ in dB scale, where $YUV\_PSNR$ = $(6.0*Y\_PSNR+U\_PSNR+V\_PSNR)/8.0$, and $Y MS$-$SSIM (dB)$ = $-10*log_{10}(1-Y MS$-$SSIM)$. The R-D curves are also shown in Fig.~\ref{fig_RD Curve}, and compared with BPG \cite{sullivan2012overview}, and VVC Test Model (VTM) \cite{vvc-vtm}.

It can be seen that our method can achieve up to 5 dB higher Y MS-SSIM than VTM at high rates. In YUV PSNR, our scheme is comparable to BPG in many cases. Better results can be obtained if the  autoregressive model in \cite{Minnen18,akbari2020generalized} is used.

The rate control of our method with only three models is very accurate. All target bitrates can be  reached by choosing a proper $\lambda$ value (which can be different from the training values), and the bitrate deviation can be made less than 1\%.

The encoding and decoding of the scheme are also very fast, only 15 and 21 secs respectively for an image of 1944x1296 pixels on CPU. The times reduce to  within 10 secs on GPU. It can be even faster if 8-bit integer models are used.

\section{Discussions}

In this proposal, we present a learned multi-frequency image compression approach that uses generalized octave convolutions. We develop a variable-rate scheme using the R-D tradeoff lagrangian parameter to modulate all the internal feature maps in the auto-encoder, which can achieve all the target bitrates for each test image by only three models. Experiments show that the proposed scheme achieves much better Y MS-SSIM than VVC and comparable results to BPG in YUV PSNR.

The proposed scheme can still be improved in many ways, for example, by incorporating other advanced entropy models. The complexity of the system can also be reduced by using methods such as model compression and optimization.

{
\bibliographystyle{ieee}
\bibliography{main}
}
\newpage
\begin{table*}
\caption{Reconstructed $PSNR\_YUV (dB)$ and $Y MS$-$SSIM (dB)$ of the specified test image jpegai03.}
\label{test03}
\center
\begin{tabular}{c|c|c|c|c|c|c|c|c}
\hline
Target bitrate (bpp)      &0.06 &0.12 & 0.25 &0.5 &0.75 &1.0 &1.5 &2.0 \\
\hline
Reached bitrate (bpp)      &0.0597 &0.1203 &0.2517 &0.5020 &0.7494 &1.0007 &1.4891 &1.9828 \\
\hline
$YUV\_PSNR (dB)$      &28.64 &31.54 &34.63 &38.09 &39.85 &41.79 &44.52 &46.24 \\
\hline
$Y MS$-$SSIM (dB)$      &13.18 &16.59 &20.26 &24.83 &27.76 &29.66 &32.57 &34.82 \\
\hline
\end{tabular}
\vspace{-0.15cm}
\end{table*}

\begin{table*}
\caption{Reconstructed $PSNR\_YUV (dB)$ and $Y MS$-$SSIM (dB)$ of the specified test image jpegai09.}
\label{test09}
\center
\begin{tabular}{c|c|c|c|c|c|c|c|c}
\hline
Target bitrate (bpp)      &0.06 &0.12 & 0.25 &0.5 &0.75 &1.0 &1.5 &2.0 \\
\hline
Reached bitrate (bpp)      &0.0599 &0.1202 &0.2499 &0.4966 &0.7494 &1.0027 &1.4976 &2.01466 \\
\hline
$YUV\_PSNR (dB)$      &20.72 &21.96 &23.37 &25.65 &26.55 &28.30 &31.14 &33.74 \\
\hline
$Y MS$-$SSIM (dB)$      &7.52 &9.86 &12.62 &16.08 &18.79 &20.92 &24.07 &27.01 \\
\hline
\end{tabular}
\vspace{-0.15cm}
\end{table*}

\begin{table*}
\caption{Reconstructed $PSNR\_YUV (dB)$ and $Y MS$-$SSIM (dB)$ of the specified test image jpegai12.}
\label{test12}
\center
\begin{tabular}{c|c|c|c|c|c|c|c|c}
\hline
Target bitrate (bpp)      &0.06 &0.12 & 0.25 &0.5 &0.75 &1.0 &1.5 &2.0 \\
\hline
Reached bitrate (bpp)      &0.0669 &0.1200 &0.2509 &0.5028 &0.7570 &0.9989 &1.5030 &2.0060 \\
\hline
$YUV\_PSNR (dB)$      &28.20 &29.94 &31.95 &34.01 &34.84 &35.96 &38.21 &40.53 \\
\hline
$Y MS$-$SSIM (dB)$      &7.69 &10.22 &13.33 &17.25 &19.85 &21.44 &24.32 &27.02 \\
\hline
\end{tabular}
\vspace{-0.15cm}
\end{table*}

\begin{table*}
\caption{Reconstructed $PSNR\_YUV (dB)$ and $Y MS$-$SSIM (dB)$ of the specified test image jpegai15.}
\label{test15}
\center
\begin{tabular}{c|c|c|c|c|c|c|c|c}
\hline
Target bitrate (bpp)      &0.06 &0.12 & 0.25 &0.5 &0.75 &1.0 &1.5 &2.0 \\
\hline
Reached bitrate (bpp)      &0.0598 &0.1203 &0.2507 &0.5006 &0.7569 &1.0076 &1.4997 &1.992 \\
\hline
$YUV\_PSNR (dB)$     &28.40 &31.21 &34.93 &38.65 &40.30 &42.14 &44.94 &46.46 \\
\hline
$Y MS$-$SSIM (dB)$      &12.50 &16.18 &20.12 &23.48 &25.69 &27.63 &30.34 &32.10 \\
\hline
\end{tabular}
\vspace{-0.15cm}
\end{table*}

\begin{table*}
\caption{Encoding and decoding time of the specified test images using CPU or a single Titan Xp GPU.}
\label{time}
\center
\begin{tabular}{c|c|c|c|c}
\hline
ImageID      &03 &09 & 12 &15 \\
\hline
Resolution (width x height)      &1944x1296 &1976x1312 & 1512x2016 &3680x2456 \\
\hline
Encoding Time on CPU/GPU (sec)      &15.33/9.50 &14.62/9.34 &15.56/9.41 &34.25/12.97 \\
\hline
Decoding Time on CPU/GPU (sec)      &21.06/8.35 &18.70/8.49 &19.50/8.74 &50.15/13.28 \\
\hline
\end{tabular}
\end{table*}

\end{document}